\documentclass[10pt,emulateapj,apj,twocolumn]{emulateapj}
\usepackage{color}
\usepackage{graphics}     
\usepackage{amsmath}
\usepackage{natbib}

\newcommand\s{\small}
\newcommand\scs{\scriptsize}
\newcommand\fns{\footnotesize}
\shorttitle{Quasars and LSS}
\shortauthors{Song et al.}
\begin{document}
\title{Quasars as a tracer of large-scale structures in the distant universe}
\author{Hyunmi Song\altaffilmark{1}, Changbom Park\altaffilmark{1},  
Heidi Lietzen\altaffilmark{2,3} and Maret Einasto\altaffilmark{4}}
\altaffiltext{1}{School of Physics, Korea Institute for Advanced Study, Heogiro 85, Seoul 130-722, Korea}
\altaffiltext{2}{Instituto de Astrofisica de Canarias, E-38205 La Laguna, Tenerife, Spain}
\altaffiltext{3}{Universidad de La Laguna, Dept. Astrofisica, E-38206 La Laguna, Tenerife, Spain}
\altaffiltext{4}{Tartu Observatory, 61602 Toravere, Estonia}
\begin{abstract}
We study the dependence of 
the number density and properties of quasars
on the background galaxy density
using the currently largest spectroscopic datasets of quasars and galaxies.
We construct a galaxy number density field smoothed over the variable smoothing scale of
between approximately 10 and $20\,h^{-1}$Mpc over the redshift range of $0.46<z<0.59$
using the Sloan Digital Sky Survey (SDSS) Data Release 12 (DR12) 
Constant MASS (CMASS) galaxies.
The quasar sample is prepared from the SDSS I/II DR7.
We examine the correlation of incidence of quasars with the large-scale background density
and dependence of quasar properties such as 
bolometric luminosity, black hole mass, and Eddington ratio on the large-scale density.
We find a monotonic correlation 
between the quasar number density and large-scale galaxy number density,
which is fitted well with 
a power law relation, $n_Q\propto\rho_G^{0.618}$.
We detect weak dependences of quasar properties on the large-scale density 
such as a positive correlation between black hole mass and density,
and a negative correlation between luminosity and density.
We discuss the possibility of using quasars as a tracer of 
large-scale structures at high redshifts,
which may be useful for studies of 
growth of structures in the high redshift universe.
\end{abstract}

\section{Introduction}
Quasars are the most luminous type of Active Galactic Nuclei (AGN)
with luminosity more than hundreds times higher than that of normal galaxies.
Thanks to such brightness, quasars can be observed
all across the universe as far as $z=7$ \citep{mortlock-etal2011,momjian-etal2014}.
Many studies of quasars have focused on 
discovering them and characterizing individual objects
to shed light on formation and evolution 
of non-linear structures at high redshifts 
\citep{turner1991,djorgovski1999,djorgovski-etal2003,djorgovski-etal2006}.
In recent years, 
large surveys such as 
the Sloan Digital Sky Survey \citep[SDSS;][]{sdss2000,gunn-etal2006}
and the 2dF QSO Redshift Survey \citep{croom-etal2004},
have accumulated large spectroscopic data sets of quasars.
This enables us to perform statistical studies of quasars
with broad scope.

Using these quasar survey data,
\citet{clowes-etal2013}, \citet{nadathur2013}, \citet{einasto-etal2014} and \citet{parkcb-etal2015}
found very large quasar groups and discussed the cosmological implications 
of the existence and properties of these extreme objects.
\citet{nadathur2013}, \citet{einasto-etal2014} and \citet{parkcb-etal2015}
pointed out that \citet{clowes-etal2013}'s cosmological interpretation of large quasar groups 
that questions the validity of the cosmological assumption of homogeneity and isotropy
is misleading.
They stressed the importance of a statistically precise analysis
to draw conclusions on cosmological implication
from existence of one or a few extreme objects in observation.
\citet{parkcb-etal2015} also emphasized that statistical comparison
with cosmological simulations must be employed as well.
The quasar survey data are also used for studies of constraining 
cosmological parameters \citep{han_park2015,risaliti_lusso2015}.
Besides the studies directly related to cosmology,
there are more studies of exploring quasar clustering properties.
\citet{einasto-etal2014} made a catalog of quasar groups 
with different linking lengths and examined their properties. 
They found that the characteristics of quasar groups
such as number density, size and richness,
identified with linking lengths varied from $20$ to $40\,h^{-1}$Mpc
are well correlated with those of galaxy superclusters. 
Therefore such quasar groups can be markers of galaxy superclusters.
As a classical way to study clustering properties,
correlation functions have been measured for quasars (AGN in general)
by many different groups
\citep{krumpe-etal2010,miyaji-etal2011,krumpe-etal2012,krumpe-etal2015,
capelluti-etal2012,allevato-etal2014,shen-etal2007c,shen-etal2009,shen-etal2013,
richardson-etal2012,eftekharzadeh-etal2015,ross-etal2009}.
They have measured two-point cross-correlation functions  (2PCCFs)
between quasars and galaxies. They have found the typical mass of 
quasar-hosting dark matter halos (DMHs)
and dependence of the mass on quasar luminosity.

Among the above-mentioned clustering studies,
\citet{shen-etal2013} measured the 2PCCFs of 
the quasars in the catalog of \citet{schneider-etal2010}
and the SDSS Constant MASS (CMASS) galaxies, 
and found that quasars at $z\sim0.5$ live in DMHs 
with mass of about $4\times10^{12}\,h^{-1}M_{\odot}$.
Even though the DMH mass gives us general information 
on the large-scale environments preferred by quasars
(the value corresponds to the scale of galaxy groups),
some environmental information, for example characterized by background density,
is not explicitly readable in the 2PCCFs typically measured in AGN clustering studies.
Such information can be preserved 
by taking a different approach from 2PCCFs,
and instead determining how quasars populate the density field traced out by galaxies.
Among the previous studies taking this approach,
\citet{lietzen-etal2009} and \citet{lietzen-etal2011}
constructed a large-scale density field, identified large-scale structures (LSSs),
and found that quasars and other AGN at $z<0.4$ are typically located
in the outskirts of galaxy superclusters.

In addition to studying structures on very large scales,
direct measures of the galaxy density field also allow us
to explore the environmental dependence of quasar properties
such as black hole mass and luminosity.
It is an approach in the reverse order of that used for the 2PCCF:
we select quasars on the basis of their environment and examine the trend
of environment-averaged properties,
while the 2PCCF studies select quasars on a quasar property
and compute the average galaxy densities around quasar subsamples.
The method used here may be
more sensitive in detecting any environmental dependences than the 2PCCF one. 
Recently, \citet{hutsemekers-etal2014} showed that quasar polarizations
are aligned to directions of the LSS to which they belong.
\citet{pelgrims_hutsemekers2015} detected large-scale alignments
of quasars polarization vectors.
There are also studies done with cosmological hydrodynamical simulations
showing that formation and evolution of galaxies
are likely to be in part driven by cosmic web 
\citep[and references therein]{laigle-etal2015}.
Based on this range of observational and theoretical work,
there is clear motivation for studying the connection between the properties
of quasars and their host LSSs.

In this paper we construct
a three-dimensional galaxy number density field
and study the occurrence of quasar as a function 
of galaxy number density at $z\sim0.5$.
A three-dimensional galaxy number density field contains
different information from the 2PCCFs,
and provides a more direct picture of LSS and matter distribution.
We will use the galaxy number density field 
to investigate the dependence of various properties of quasars,
such as luminosity, black hole mass and Eddington ratio,
on the background galaxy density.
It should be noted that in this paper we focus on how quasar activity is related to
the local galaxy density traced by massive galaxies, 
rather than to the underlying local matter density.
We adopt a flat $\Lambda$CDM cosmology with 
$\Omega_M=0.27$, $\Omega_\Lambda=0.73$ and $h=0.7$, 
and comoving distances are used throughout the paper.

\section{Data}
We use the fifth edition of the SDSS quasar catalog of \citet{schneider-etal2010}
from SDSS DR7 which is a compilation of quasars
observed by the SDSS-I/II quasar survey.
The catalog contains 105783 spectroscopically confirmed quasars 
over a wide redshift range of $0.065<z<5.46$
in the area covering approximately $9380\,\textrm{deg}^2$ of the sky.
They are brighter than the $i$-band absolute magnitude of $M_i=-22.0$ 
($M_i$ is galactic extinction-corrected and K-corrected to $z=2$
in a cosmology with $H_0=70\,$km/s/Mpc, 
$\Omega_m=0.3$ and $\Omega_\Lambda=0.7$)
and have at least one broad emission line 
with Full Width at Half Maximum (FWHM) larger than $1000\,$km/s
or interesting/complex absorption features.
They have apparent $i$-magnitude of $14.86<i<22.36$, 
where the bright limit comes from the maximum
brightness limit of the target selection on quasar candidates 
to avoid saturation and cross-talk in the spectra.
The catalog does not include several classes of AGN 
such as Type II quasars, Seyfert galaxies and BL Lacertae objects.
For the details of target selection process for spectroscopic observation
and quasar confirmation process, 
see \citet{richards-etal2002} and \citet{schneider-etal2010}.
The original survey of the SDSS quasars turned out to be non-uniform
in the selection of targets.
To remedy this problem \citet{shen-etal2011} 
evaluated the SDSS target selection
and provided a so-called uniform flag in their catalog.
The quasar sample is expected to be statistically uniform
when only the quasars with the uniform flag equal to 1 are selected
\citep{richards-etal2002,richards-etal2006,shen-etal2007c}.

To construct a large-scale galaxy density field 
we use the latest CMASS galaxy catalog \citep[DR12v4;][]{alam-etal2015}
of SDSS-III\citep{eisenstein-etal2011}
Baryonic Oscillation Spectroscopic Survey 
\citep[BOSS;][]{dawson-etal2013,smee-etal2013}, 
which contains 621849 galaxies
mostly in the redshift range of $0.4<z<0.8$. 
The CMASS sample is designed to have massive galaxies 
to detect Baryonic Acoustic Oscillation (BAO) around $z\sim0.5$
through a set of flux and color selection criteria.
For more details about the CMASS target selection criteria,
see \citet{eisenstein-etal2001}, \citet{cannon-etal2006}
and \citet{bolton-etal2012}.
Those massive BOSS CMASS galaxies are an optimal tracer of
the large-scale matter distribution.

The stellar masses of CMASS galaxies are above $10^{11.2-11.3}M_\odot$ 
\citep{chen-etal2012,maraston-etal2013}. We use the group catalog of the SDSS main sample by
\citet{tempel-etal2014} as a comparison to determine what fraction of the high-mass galaxies
are central galaxies of their groups.
We find that $87\%$ of the galaxies above $10^{11.3}M_\odot$ in the distance bin
from $120$ to $340\,h^{-1}$Mpc are the first-rank galaxies (most luminous) of their groups \citep[see also][]{leitzen-etal2016}.
In that sense, the CMASS sample does not suffer much from peculiar motions,
thus we do not need to worry about contamination 
by galaxy peculiar motion in the galaxy density estimate. 
Even for $13\%$ of non-central galaxies,
the galaxy density we calculate is a value smoothed over a larger scale
($\sim15.2h^{-1}$Mpc; see Section 3.2) than a typical uncertainty level
due to peculiar motion of a galaxy cluster ($<5h^{-1}$Mpc in comoving scale at $z\sim0.5$).

The BOSS CMASS sample suffers from various systematics
of missing galaxies due to fiber collisions, poor or failed observations, and so on.
In calculating the galaxy number density we will use a weight given to each galaxy
to take the systematic effects into account.
More details on the weight will be given in the next section.

To make a statistically homogeneous sample of density tracers
we apply a cut to the galaxy sample in $i$-band absolute magnitude versus redshift space
so that the comoving number density of galaxies becomes roughly constant of redshift.
The cut is obtained via fitting with a functional form of
\begin{equation}
   M_{i,\textrm{cut}} = 0.3 \left(\frac{\pi}{2}-\textrm{atan}(41.9(z-0.459))\right) - 22.6.
\label{eq_mag_cut}
\end{equation}
The resulting sample has the mean galaxy separation of $17.1\,h^{-1}$Mpc,
which is the cube root of the typical volume occupied by each galaxy.
We limit our sample within the redshift range of $0.45<z<0.61$,
where the cut does not change drastically (meaning survey completeness guaranteed)
and as many galaxies as possible can be kept while satisfying the constant number density condition.

\begin{figure*}
\centering
\includegraphics[width=\textwidth]{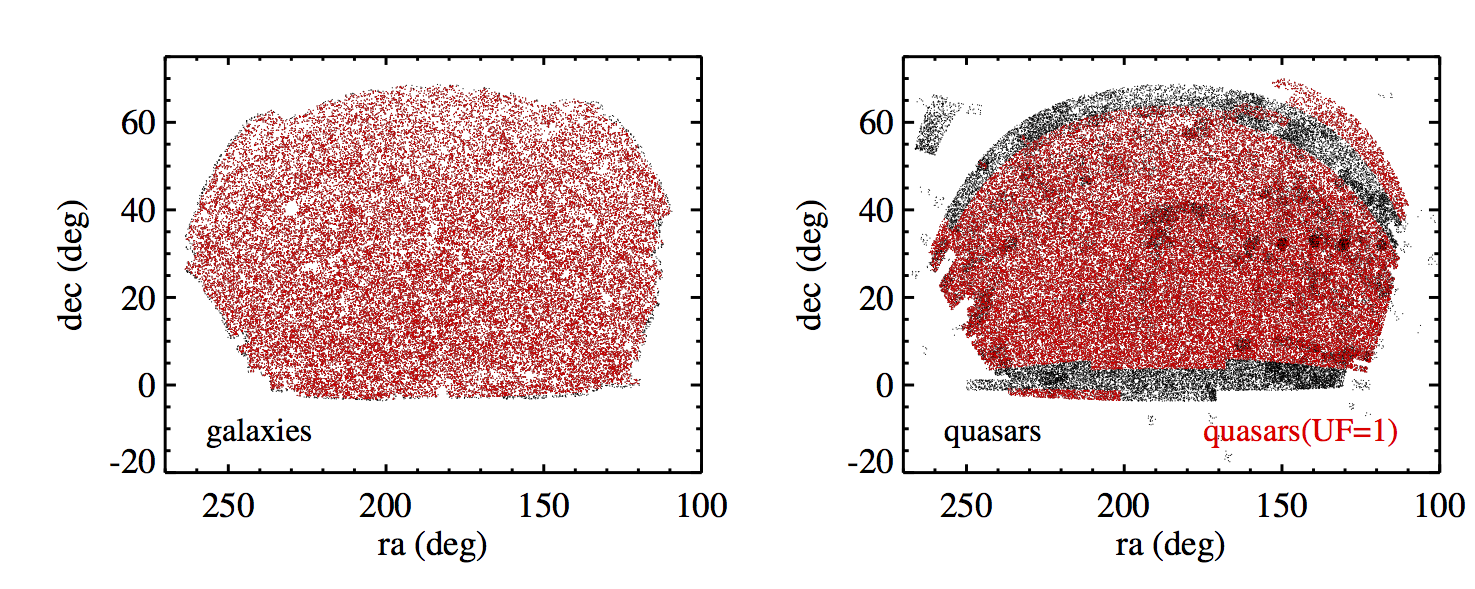}
\caption{
Distributions of SDSS-III DR12v4 CMASS galaxies (left) 
and SDSS DR7 quasars (right) 
in the northern galactic cap.
Only about 10\,$\%$ of galaxies are shown for clarity.
Black dots in both panel represent the full samples.
Galaxies within our sample boundaries 
used to calculate the density field are colored in red.
Quasars flagged with $\textrm{UF}=1$ by \citet{shen-etal2011} 
are over-plotted in red in the right panel,
and our quasar sample is limited in the middle, largest patch.
In our analysis we use the overlapped region of the galaxy and quasar samples.
}
\label{fig_sky_distr_all}
\end{figure*}

\begin{figure*}
\centering
\includegraphics[width=0.8\textwidth]{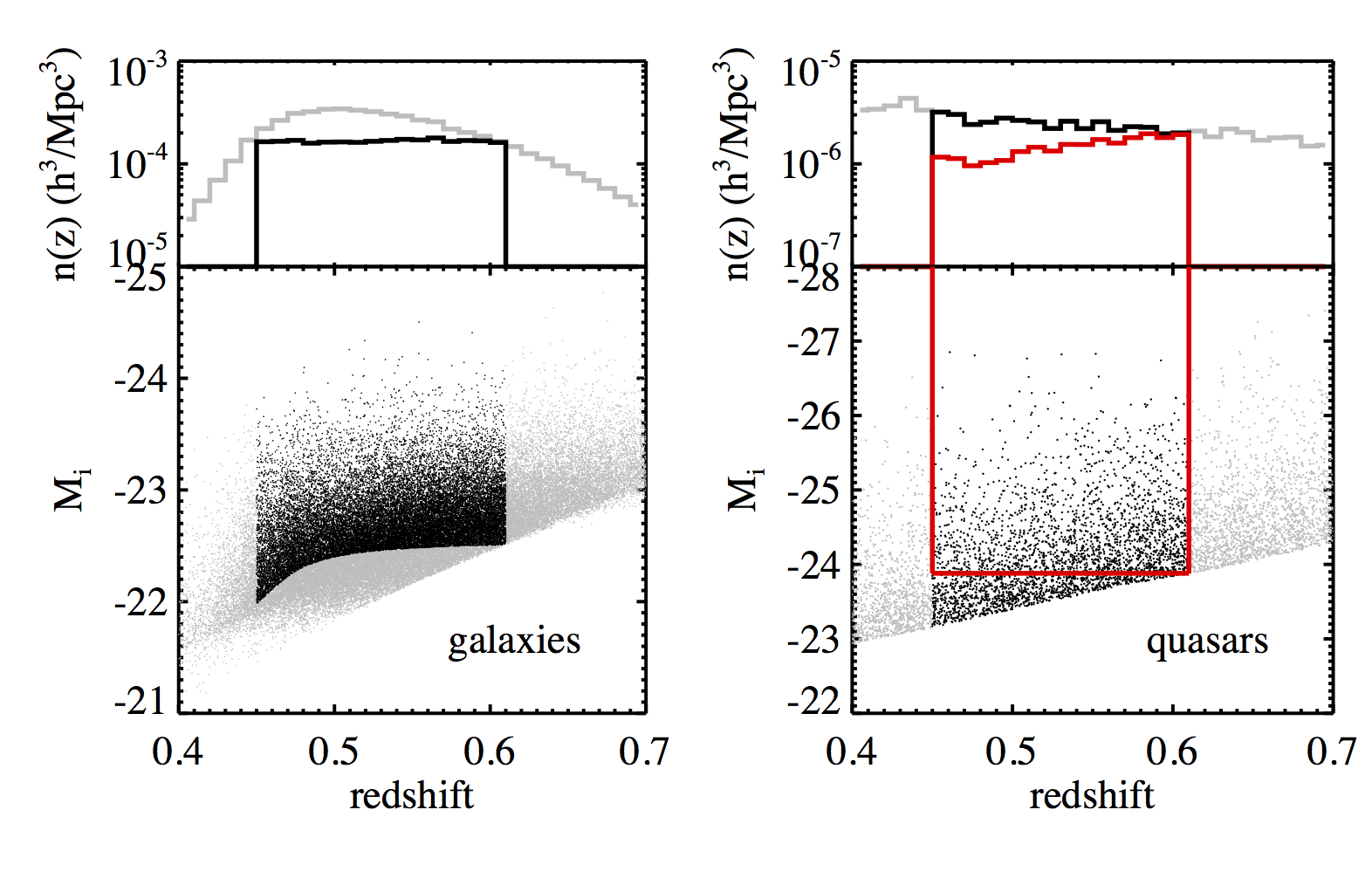}
\caption{
Distributions of the SDSS-III CMASS galaxies (left) 
and SDSS DR7 quasars (right) 
in the $i$-band absolute magnitude in redshift space
with (comoving) number density histograms as a function of redshift above.
Black dots represent those selected for the analysis of this paper.
The absolute magnitude cut 
given by Equation (\ref{eq_mag_cut}) is applied
to the galaxy sample to make a constant number density sample 
with the mean galaxy separation of $17.1\,h^{-1}$Mpc
over the redshift range $0.45<z<0.61$.
Only about $10\,\%$ of galaxies are plotted.
All quasars in the redshift range of $0.45<z<0.61$ are
shown in black.
When we study the redshift evolution of quasar properties,
a constant absolute magnitude cut of $M_i<-23.9$,
shown with the red line, is used for quasars.
}
\label{fig_z_absmag}
\end{figure*}

Figure \ref{fig_sky_distr_all} shows the distributions of galaxies (left)
and quasars (right) in the northern galactic cap (NGC).
To reduce the effects of sample boundaries in the calculation of local density
some of the jagged boundaries of the CMASS galaxy distribution are cut off,
which removes about $2\%$ of CMASS galaxies.
Galaxies located within the newly defined boundaries are shown in red
and they are used to calculated the density field.
In the right panel quasars having the uniform flag equal to 1 
and also within a cleaned boundaries are shown in red.
It is seen that the sample of quasars with the uniform flag of 1
is divided into three contiguous regions. We use the subsample
in the middle that has the largest volume. Our analyses are limited within
the region where the galaxy and quasar samples overlap each other.
Figure \ref{fig_z_absmag} shows the selected galaxies and quasars
for our analysis in the plane of $i$-band absolute magnitude and redshift 
with black dots. The gray dots represent those in the original samples
within the NGC and cleaner boundaries.

\section{Galaxy density field}
\subsection{Calculating the smooth galaxy density field}
We use a smooth number density field of the CMASS galaxies
to examine how quasars are distributed relative to the LSSs of galaxies.
We apply the Spline-kernel smoothing with variable kernel size
to the sample of CMASS galaxies prepared as described in the previous section.
We embed our sample within a large cuboid and calculate the local density
at the center of each cell, which is a cube with side length of $5\,h^{-1}$Mpc,
following the widely used method in smoothed particle hydrodynamics (SPH)
\citep{monaghan1992} among many other non-parametric regression estimations 
\citep{jennen-steinmetz_gasser1988}.
The local galaxy number density at each cell is formulated as
\begin{equation}
  \rho = \sum_{i=1}^{N_\textrm{\fns nn}} W(r_i,h_\textrm{\s spl})
\end{equation}
where $W(r_i,h_\textrm{\s spl})$ is a Spline function kernel.
$r_i$ is the distance between the cell center and its $i$-th nearest neighboring galaxy, and
$h_\textrm{\s spl}$ is a smoothing length defined as $r_{N_\textrm{\fns nn}}/2$.
$N_\textrm{\fns nn}$ is number of nearest neighbors used for density calculation
and we choose 20.
The choice of 20 is made based on the test of which value of $N_\textrm{\fns nn}$ efficiently
reproduces a uniform distribution 
with less than $1\%$ error (see Appendix for more details).
$r_{20}$ varies from a location to another reflecting the density fluctuations.
The Spline function kernel we adopt here is from \citet{monaghan_lattanzio1985}
\footnote{We performed a test of comparing densities calculated 
with the two different Spline function kernels, $W_3$ and $W_4$ of \citet{monaghan_lattanzio1985}.
We showed that the two densities agree well with each other and the main results of this paper
are robust to the choice of a Spline function kernel.}:
\begin{equation}
  W(r_i,h_\textrm{\s spl})
     =\frac{1}{\pi h_\textrm{\fns spl}^3} 
      \begin{cases}
             1-\frac{3}{2}q_i^2 + \frac{3}{4}q_i^3 \,; \quad q_i\le1\\
             \frac{1}{4}(2-q_i)^3 \,; \quad 1\le q_i\le2\\
             0 \,; \qquad\qquad \text{otherwise}
      \end{cases}
\end{equation}
where $q_i=r_i/h_\textrm{\s spl}$.
This function satisfies accuracy, smoothness and computational efficiency requirements
for the interpolating kernels
with its compact support and continuous second derivative.
Please see \citet{monaghan_lattanzio1985} and \citet{monaghan1992} for more details.

As already mentioned in Section 2, 
some galaxies are missing from the observation 
and data processing and analysis for various reasons:
(1) targets that have spectroscopic redshifts in the literature 
(denoted as {\it known});
(2) targets that have a target of different type (i.e. QSO) 
getting a fiber allocation already within $62''$ (denoted as {\it missed});
(3) targets that have another CMASS target within $62''$ 
(denoted as {\it close pair} or shortly {\it cp});
(4) targets that were assigned fibers but after spectroscopic observation, 
were revealed as stars (denoted as {\it star});
(5) targets that were assigned fibers, observed,
but did not give a reliable redshift through the pipeline 
for various reasons (denoted as {\it fail}). 
So (2) and (3) are the cases of fiber allocation failure, 
and (4) and (5) are the cases of spectroscopic observation failure.

The completeness of the CMASS galaxy catalog 
at a location of the sky can be quantified by 
\begin{equation}
  C_\textrm{\s sample}
  =\frac{ N_\textrm{\s gal}+N_\textrm{\s known}+N_\textrm{\s cp} }
   { N_\textrm{\s gal}+N_\textrm{\s known}+N_\textrm{\s cp}+N_\textrm{\s missed}+N_\textrm{\s fail} }
  \label{eq_c_sample}
\end{equation}
where $N_\textrm{\s gal}$ is the number of galaxies in the CMASS galaxy catalog
which were assigned fibers and successfully observed with spectroscopy.
Here we assume that all of objects which are missed and failed are CMASS galaxies.
$C_\textrm{\s sample}$ would be defined in each sector,
which is an area covered by a unique set of spectroscopic tiles
where the observing conditions are the same.
But it is not easy to find all $N$'s used in Equation \ref{eq_c_sample}
in each sector. Rather what is given in the catalog sector-by-sector
is a similar quantity,
\begin{equation}
  C_\textrm{\fns BOSS}=\frac{N_\textrm{\s obs}+N_\textrm{\s cp}}{N_\textrm{\s targ}-N_\textrm{\s known}}
  \label{eq_c_boss}
\end{equation}
where $N_\textrm{\s obs}=N_\textrm{\s star}+N_\textrm{\s gal}+N_\textrm{\s fail}$
and $N_\textrm{\s targ}=N_\textrm{\s obs}+N_\textrm{\s cp}+N_\textrm{\s known}+N_\textrm{\s missed}$.
It tells how well spectroscopic observation is done.

Since $N_\textrm{\s gal}$ dominates over others, 
there is only small difference between $C_\textrm{\s sample}$ and $C_\textmd{\fns BOSS}$.
To compare $C_\textrm{\fns sample}$ and $C_\textrm{\scs BOSS}$
we calculate them using the information given in 
table 1 of \citet{anderson-etal2014},
and find that for DR11 CMASS galaxies in northern hemisphere, 
$C_\textrm{\fns sample}=0.9999\,C_\textrm{\scs BOSS}$ 
in the survey area as a whole. 
To be more precise, we need to compare them in every sector. 
But the spatial variation of $C_\textrm{\scs BOSS}$ 
across different sectors is small and
the CMASS targets which contribute to the sample incompleteness 
should be distributed randomly in the sky. 
So we expect the global relation of 
$C_\textrm{\fns sample}=0.9999\,C_\textrm{\scs BOSS}$ 
could be also satisfied locally.
Therefore, for convenience, 
we adopt $C_\textrm{\fns BOSS}$ for $C_\textrm{\s sample}$.

We correct the local density estimates for the effects
of missing galaxies by using $C_\textrm{\fns BOSS}$ and the weight
\begin{equation}
  \omega_\textrm{\s tot}=(\omega_\textrm{\s cp} + \omega_\textrm{\s rf} -1)\,\omega_\textrm{\s sys}.
\end{equation}
$\omega_\textrm{\s cp}$ and $\omega_\textrm{\s rf}$ are weights for fiber collisions of close pairs
and redshift failures respectively. Both of them are assigned one by default,
but up-weighted for galaxies that have colliding targets of same type (galaxy)
and objects with failed redshift as nearest neighbors respectively.
$\omega_\textrm{\s sys}$ takes into account the spurious fluctuations
in galaxy distribution caused by distribution of stars,
galactic extinction, seeing, airmass and sky background (imaging systematics).
See \citet{anderson-etal2014} and \citet{ross-etal2012} 
for more information about the weights and the completeness.
The resulting total weight to each galaxy is
$\omega_\textrm{\s fin}=\omega_\textrm{\s tot}/C_\textrm{\fns BOSS}$.
When the local density is calculated at each cell,
instead of fixing $N_\textrm{\s nn}$ to 20,
we search for the largest $N_\textrm{\s nn}$ that satisfies
\begin{equation}
  \sum_{i=1}^{N_\textrm{\fns nn}} \omega_{i,\textrm{\s fin}} < 21.
  \label{eq_new_ns}
\end{equation}
Then the density is calculated from
\begin{equation}
  \rho_{\s 20} 
  = \frac{1}{\pi h_\textrm{\fns spl}^{3}} 
    \sum_{i=1}^{N_\textrm{\fns nn}} W(r_i,h_\textrm{\s spl})\,\omega_{i,\textrm{\s fin}}.
  \label{eq_new_rho}
\end{equation}
For the analysis of dependence of quasar properties on background galaxy density,
we calculate Equation (\ref{eq_new_rho}) at the position of each quasar
instead of assigning the density at the center of a cell, 
because the quasars are not always located at their cell center.

\subsection{Dealing with boundary effects}
When the smooth local density is estimated,
care should be taken on the effects of sample boundaries
made by the survey definition and screened regions.
At each cell we count the number of `active' cells
within the distance of $2h_\textrm{\s spl}$
that belong to the sample volume,
and calculate the ratio of the volume occupied by these
`active' cells to the volume of the sphere of radius $2h_\textrm{\s spl}$.
The density estimate is corrected by the factor given by 
the inverse of the ratio.
When the ratio is less than 0.8, we discard the cell (about $4\%$) 
to guarantee high quality density estimation with insignificant shot noise effects.

While dealing with the boundary effects,
cells having low density values are preferentially discarded,
thus it causes artificial, secondary boundary effects.
It especially results in fake shortages of low-density-cells
near the both ends of the redshift range we chose ($0.45<z<0.61$).
Since we are going to analyze redshift dependence
of quasar properties with density,
we exclude these artifacts near the redshift boundary
by limiting redshift range to be $0.46<z<0.59$.

\begin{figure}
\centering
\includegraphics[width=\columnwidth]{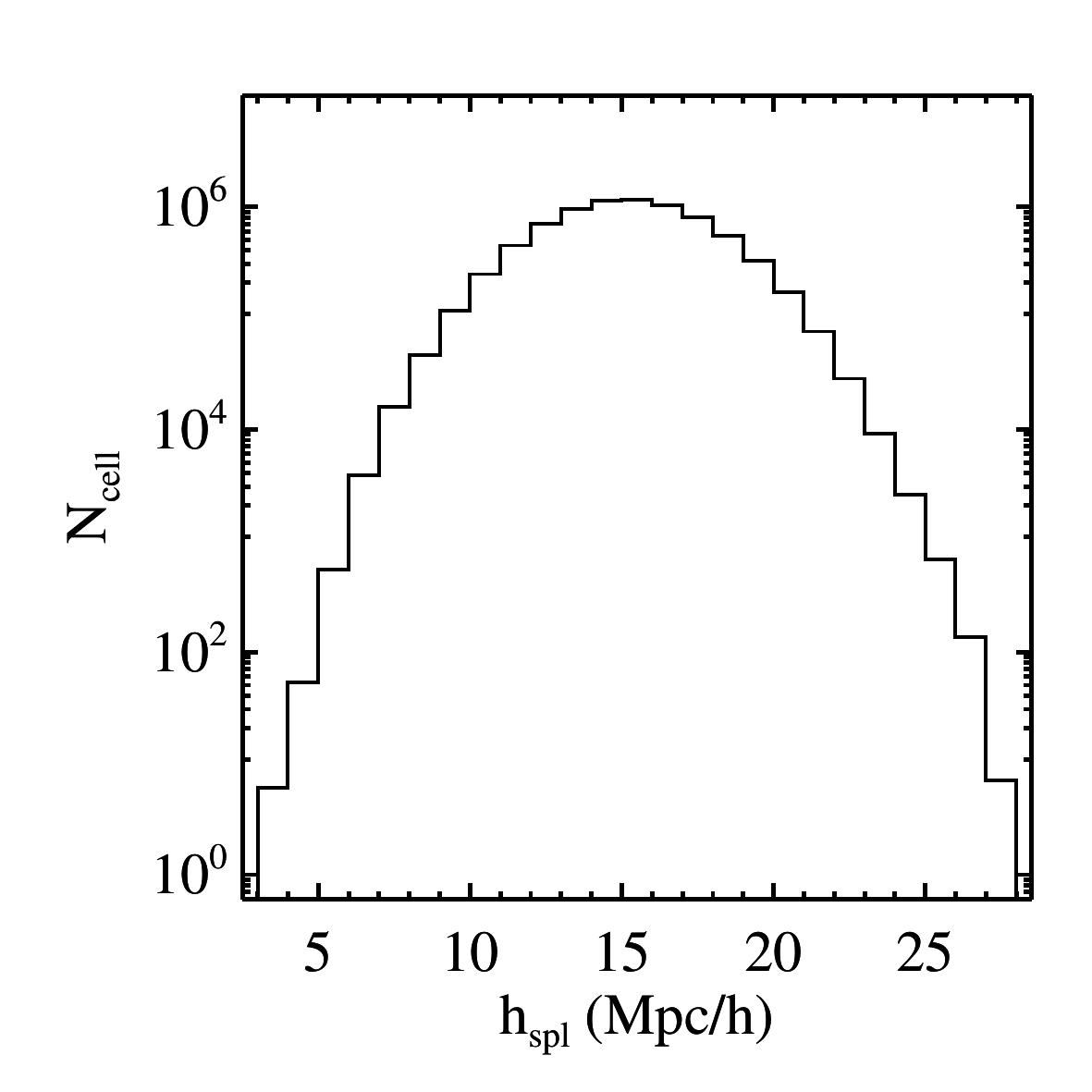}
\caption{
Frequency of the smoothing kernel size
$h_\textrm{\s spl}$ used to calculate
the local galaxy number density.
}
\label{fig_hspl}
\end{figure}
Figure \ref{fig_hspl} shows the distribution of $h_\textrm{\s spl}$ 
at the cells selected after the boundary effects are taken into account.
$h_\textrm{\s spl}$ typically has a value of $15.2\,h^{-1}$Mpc 
and a dispersion of about $2.65\,h^{-1}$Mpc.
\begin{figure*}
\centering
\includegraphics[width=\textwidth]{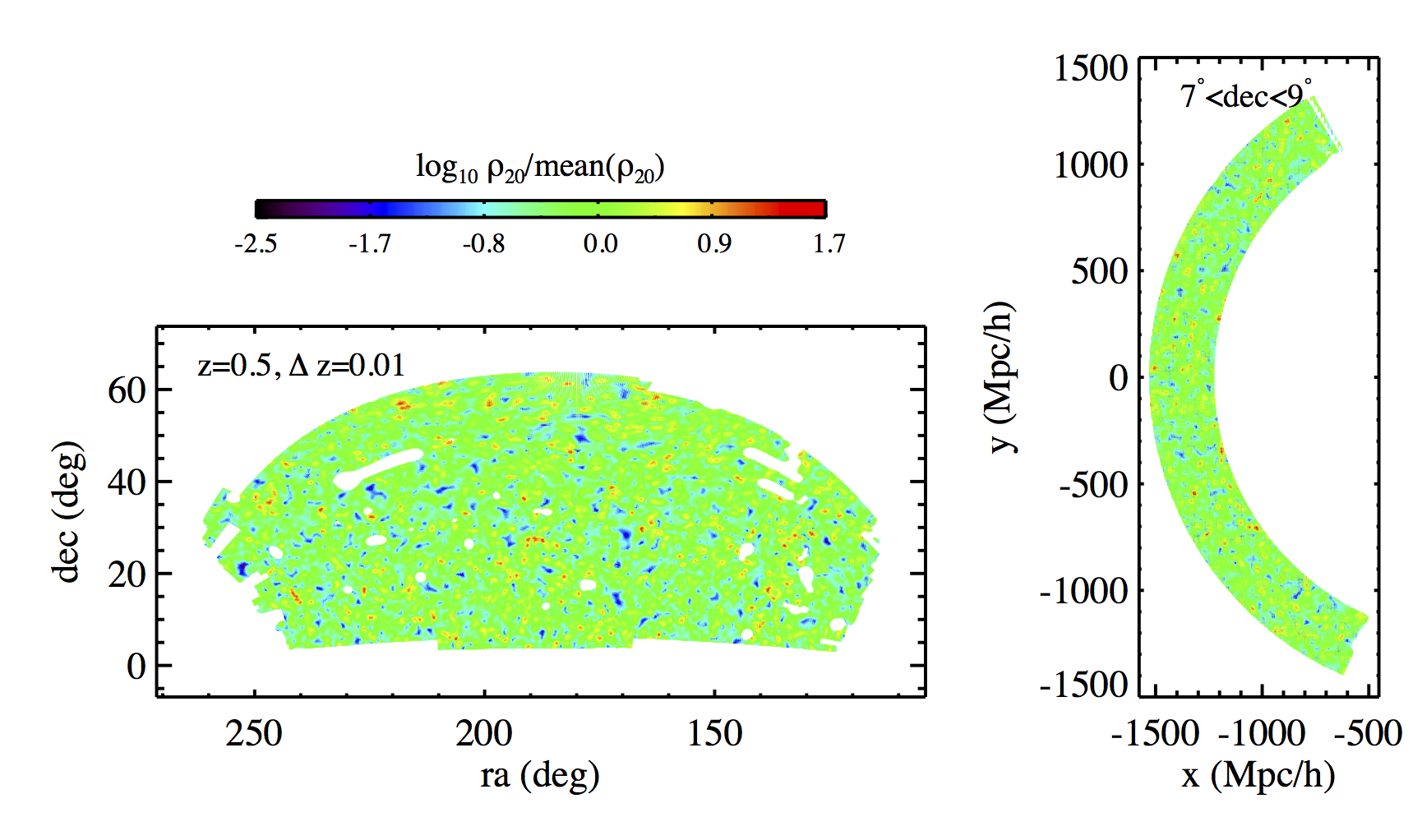}
\caption{
(Left) The galaxy density map in a thin slice 
at a median redshift of 0.5 with width of 0.01
projected on the plane of RA and Dec.
(Right) The density map in a thin slice with 
$7^{\circ}<\textrm{Dec}<9^{\circ}$ 
projected into two-dimensional comoving Cartesian space.
}
\label{fig_rho20_map}
\end{figure*}
Figure \ref{fig_rho20_map} shows the galaxy number density fields
in two thin slices. The map on the left is a slice at $z=0.5$ 
with width of $\Delta z=0.01$
projected on the sky, and the right one is a slice with 
$7^{\circ}<\delta<9^{\circ}$ projected in the x-y plane
of the equatorial coordinate system.

\section{Results}
\subsection{Quasar phenomenon versus galaxy density}
In this section we will present our main results
on the probability of finding quasars
when the local galaxy number density is given.
\begin{figure*}
\centering
\includegraphics[width=\textwidth]{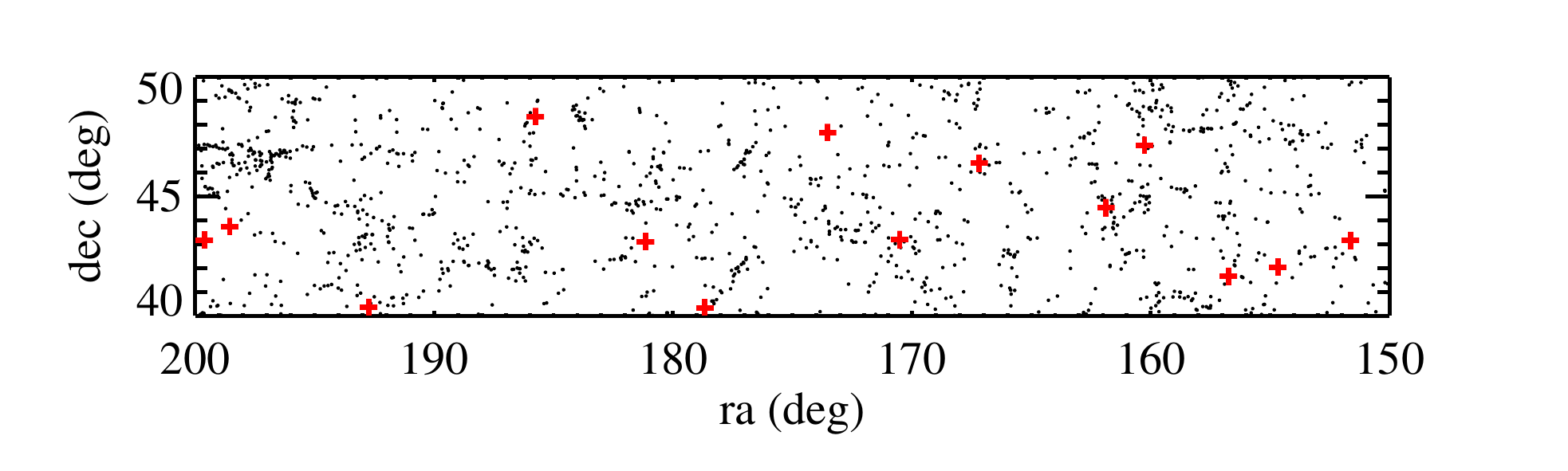}
\caption{
Quasars (red crosses) are shown
on top of galaxy (black dots) distribution 
in a stripe at a median redshift of 0.604 and with width of 0.01.
}
\label{fig_sky_distr}
\end{figure*}
Figure \ref{fig_sky_distr} compares the distribution of galaxies
(black dots) with that of quasars (red crosses)
in a thin stripe at $z=0.604$ with width of $\Delta z=0.01$.
The LSSs are not well-traced by them
because both samples are sparse.
Nevertheless, it can be noticed that two distributions are correlated.
The correlation is not strong as evidenced by a few quasars
located at empty regions and by high-density regions with no quasars.

\begin{figure}
\centering
\includegraphics[width=\columnwidth]{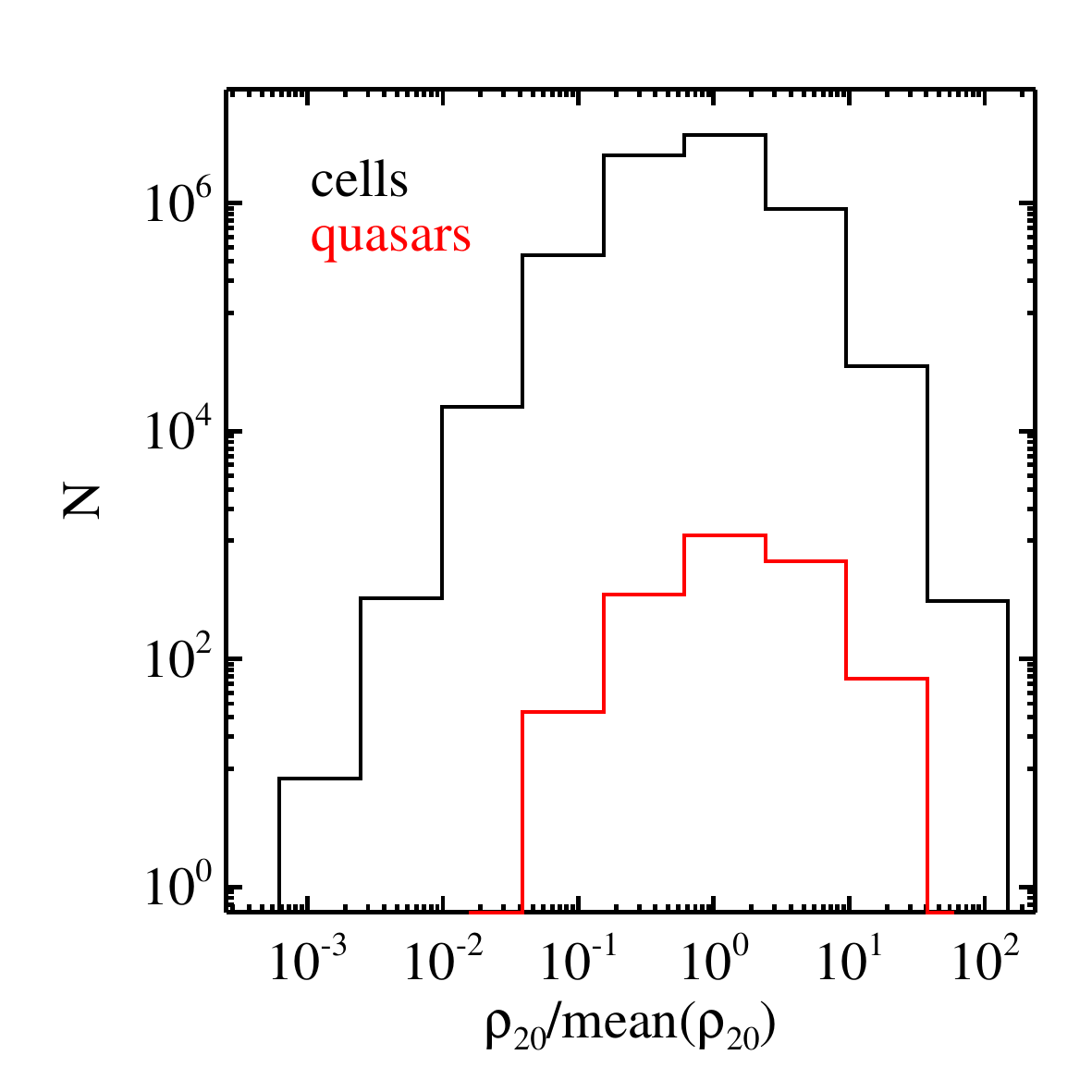}
\caption{
The numbers of cells (black) and quasars (red)
whose local density belongs to each density bin.
}
\label{fig_rho20}
\end{figure}
Figure \ref{fig_rho20} shows the number of density cells (black histogram)
having the local galaxy density given by the x-axis.
The red histogram is the number of quasars whose local density belongs
to each density bin.
The quasar number density is calculated from 
$N_\textrm{\s quasar}/N_\textrm{\s cell}V_\textrm{\s cell}$
in each bin of galaxy number density.
$N_\textrm{\s cell}$ is the number of cells in a given galaxy density bin
and $N_\textrm{\s quasar}$ is the number of quasars contained within those cells.
$V_\textrm{\s cell}$ is the volume of a cell, $125\,(h^{-1}\textrm{Mpc})^3$.
\begin{figure}
\centering
\includegraphics[width=\columnwidth]{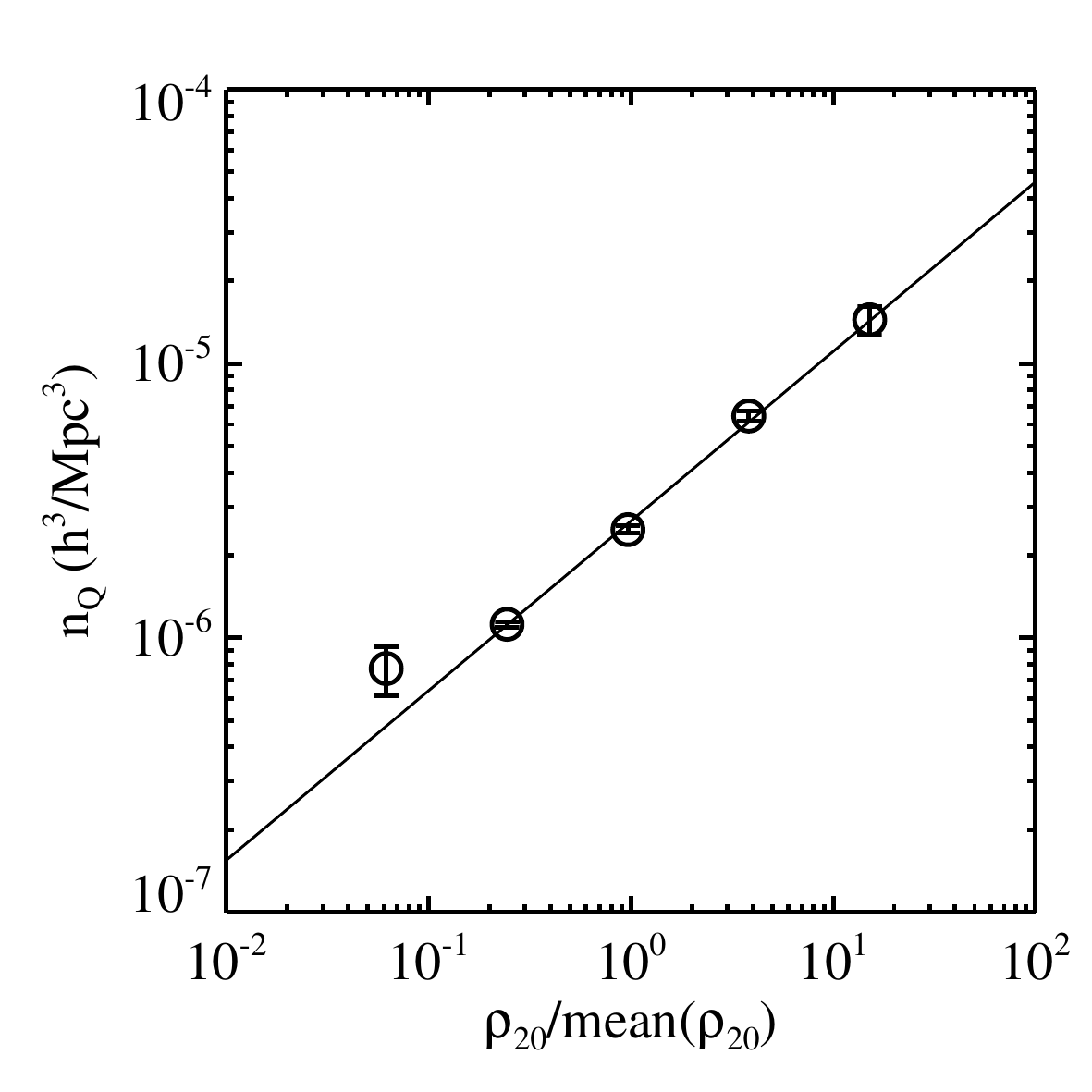}
\caption{
Quasar number density as a function of galaxy density.
Error bars are calculated from the subsample-to-subsample variation.
The solid line is the best linear fit.
The mean quasar number density is $2.45\times10^{-6}\,(h^{-1}\textrm{Mpc})^{-3}$.
}
\label{fig_rho20_nQ}
\end{figure}
Figure \ref{fig_rho20_nQ} shows that
the quasar number density is monotonically proportional 
to the background galaxy number density
over the whole density range from 
$\sim5\times10^{-2}$ to $\sim2\times10^{1}$
times the mean density.
The probability of finding quasars increases by more than
an order of magnitude over this interval.
The error bars are estimated with eight subsamples
having one-eighth of the survey area on the sky
\footnote{The error of quasar number density estimation 
at $i$-th background density bin is calculated as 
$\sigma_i^2=\frac{1}{N(N-1)}\sum_{k=1}^{N}(x_{i,k}-\bar{x_i})^2$
where $N$ is the number of subsamples, 
$x_{i,k}$ is the measurement from subsample $k$,
and $\bar{x_i}$ is the mean of the measurements from all subsamples.
}.
The best linear fit shown by a solid line is 
\begin{equation}
  \textrm{log}\,n_Q/(h^3/\textrm{Mpc}^3) = \alpha + \beta\,\textrm{log}\,\rho_{20}/\bar{\rho_{20}}
  \label{eq_fit}
\end{equation}
with $\alpha=-5.57\pm0.02$ and $\beta=0.618\pm0.034$.
The slope $\beta$ is smaller than 1,
meaning that quasar density changes slower than galaxy density does.
There is a hint for weaker correlation in low-density region.
Our result is qualitatively consistent with \citet{shen-etal2013}'s
finding that clustering of quasars is positively correlated with
that of CMASS galaxies.
To perform a quantitative comparison, we need an auto-CF of CMASS galaxies
additionally and also the conversion between 
two-dimensional \citep{shen-etal2013} and three-dimensional (this paper) quantities.
Also with the 2PCCF of \citet{shen-etal2013} one can explore only down to regions
of mean density, which is a smaller range than what we consider here.
So, we simply note 
that the gentle slope of the relation between quasar and galaxy densities we found
can be also inferred from Figure 5 of \citet{shen-etal2013}.
To draw this point, we use the relation $n(r)=n_b(1+\xi_{ab}(r))$ 
where $n(r)$ is the mean density of $b$-type objects {\it at distance $r$} from $a$-type,
$n_b$ is the mean number density of $b$-type objects, and $\xi_{ab}$ is the CCF of $a$-type and $b$-type objects
\citep[][Equation (44.4)]{peebles1980}.
We apply this equation to our case by setting $a=\textrm{galaxy}$, $b=\textrm{quasar}$ for for quasar-galaxy CCF 
and $a=b=\textrm{galaxy}$ for galaxy ACF. 
The galaxy ACF decreases more rapidly than the quasar-galaxy CCF \citep[Figure 5 of][]{shen-etal2013}.
Therefore, as a function of increasing distance $r$ from a galaxy,
the mean density of galaxies decreases rapidly with $r$ than the mean density of quasars does.

We also examine if the relation 
between quasar density and galaxy density
changes with redshift or not.
In this particular study 
we use a quasar sample defined by a uniform absolute magnitude cut of $M_i$.
We divide the sample volume into 3 sub-volumes with redshift ranges of
$0.46<z<0.52$ (low-$z$), $0.52<z<0.56$ (intermediate-$z$) and $0.56<z<0.59$ (high-$z$).
The 3 sub-volumes contain the same number of quasars.
\begin{figure}
\centering
\includegraphics[width=\columnwidth]{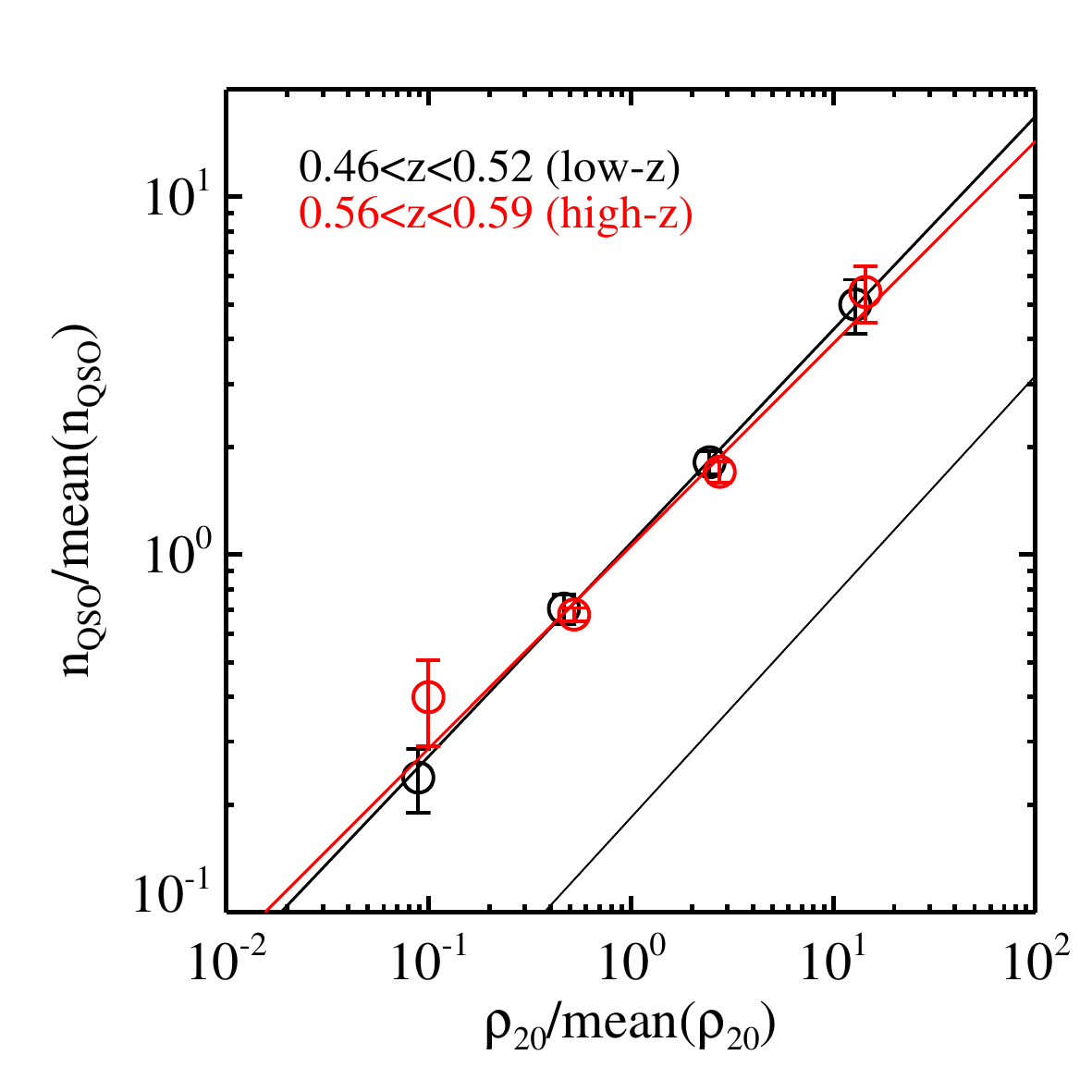}
\caption{
Same as \ref{fig_rho20_nQ},
quasar density is calculated as a function of
galaxy density, but separately for two redshift-subsamples,
low (black) and high (red) redshift subsamples.
Solid lines in black and red drawn together with circles 
are best fits respectively.
The thinner solid line in right bottom side of the figure
having the same slope with the best fit in Figure \ref{fig_rho20_nQ} 
(Equation (\ref{eq_fit})) is drawn for comparison.
}
\label{fig_rho20_nQ_red}
\end{figure}
In Figure \ref{fig_rho20_nQ_red}
the relations found in the low-$z$ and high-$z$ subsamples are compared.
Since the mean number density of quasars brighter than $M_i$
is different for those subsamples,
the quasar number density is normalized by its mean value in each subsample.
Best fits for the both cases are calculated and drawn in the figure,
and as comparison a solid line having the same slope 
of Equation (\ref{eq_fit}) is also added.
It is found that the relations are consistent with each other 
and redshift evolution of the relation is not detected.
Figure 11 of \citet{shen-etal2013} which summarizes the results
of previous 2PCF studies for quasars and galaxy, showing no change in biases
of quasar and galaxy in the redshift range similar to ours.
It means that there is no redshift evolution of quasar density-galaxy density
relation in the given redshift range.
Since there is no galaxy bias measurement at a higher redshift to compare with quasar bias,
it is difficult to say if the density relation will change or not in a wider redshift range.
Further study is needed over a wider redshift range 
(see 4th paragraph in Section 5 for more discussion).

\subsection{Quasar properties as a function of galaxy density}
We examine how quasar properties such as 
bolometric luminosity ($L_\textrm{\s bol}$),
black hole mass ($M_\textrm{\fns BH,vir}$)
Eddington ratio ($\lambda$) 
and differential color ($\Delta (g-i)$)
depend on background galaxy density. 
As mentioned in Section 3.1,
the galaxy density is calculated in the same way as described in Section 3,
but at the position of each quasar.
$M_\textrm{\fns BH,vir}$ is the 
adopted fiducial virial black hole mass calculated by \citet{shen-etal2011}.
$\Delta (g-i)$ is given by
$(g-i)_\textrm{\fns QSO} - <(g-i)>_\textrm{\s redshift}$ 
where $<(g-i)>_\textrm{\s redshift}$ is a representative color
at the redshift of a quasar
which is derived from a well-known tight color-redshift
correlation of quasars 
\citep{richards-etal2001,richards-etal2003,schneider-etal2007}.
So $\Delta (g-i)$ can tell whether or not an object 
has a redder or bluer continuum than the typical quasar at the same redshift.
All the quasar property parameters are given by
\citet{schneider-etal2010} and \citet{shen-etal2011}.
We exclude some quasars with missing information from our analysis.
These quasars are mainly at high redshift or intrinsically faint,
which may have affected the quality of observation.

We bring one more quantity, $R_\textrm{\fns FeII}$, which is 
the ratio between equivalent widths of the Fe line 
within $4435-4685\,\AA$ ($EW_\textrm{\fns FeII}$)
and broad H$\beta$ line ($EW_{\textrm{\fns H}\s\beta}$).
It is known as a good probe of black hole mass \citep{shen_ho2014}.
The black hole mass provided by \citet{shen-etal2011}
is a derived quantity determined from a number of measurements 
and based on some assumptions, while $R_\textrm{\fns FeII}$
is a single measured quantity 
and so will suffer from fewer systematic and statistical uncertainties
\citep{shen-etal2011,shen_ho2014}.

\begin{figure}
\centering
\includegraphics[width=\columnwidth]{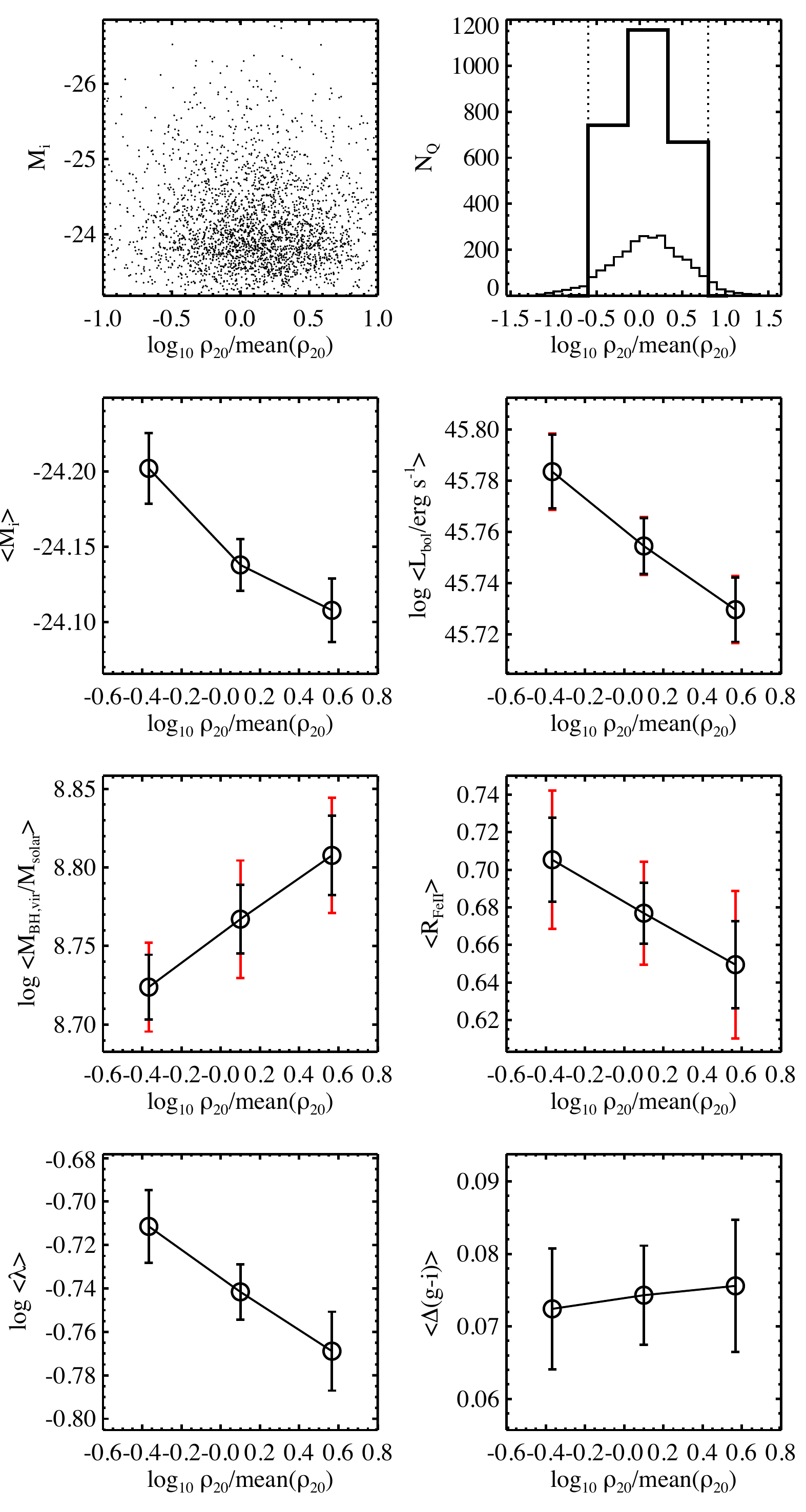}
\caption{
(Top, left) Scatter plot of quasars 
in the plane of $M_i$ and local density.
(Top, right) Numbers of quasars in three bins of local density.
The three bins have the same width in log scale.
(Rest) Averaged value of each quasar property in each bin
is shown with error bars.
Black error bars are the standard error of the mean
and red ones are the measurement error.
The quasar properties considered here are
absolute $i$-band magnitude ($M_i$),
bolometric luminosity ($L_\textrm{\s bol}$), 
black hole mass ($M_\textrm{\fns BH,vir}$), 
equivalent width ratio of Fe line and broad H$\beta$ line 
($R_\textrm{\fns FeII}$),
Eddington ratio ($\lambda$), 
and differential color ($\Delta(g-i)$). 
}
\label{fig_rho20_Qproperties}
\end{figure}

We show the scatter plot of $M_i$ 
of all quasars with respect to local galaxy density
in the top left panel of Figure \ref{fig_rho20_Qproperties}.
We inspect the dependence of quasar properties
across three density bins where statistics is high 
as shown by the number of quasars in the top right panel.
The error bars in black are the standard error of the mean
and red error bars above those in black are measurement errors from \citet{shen-etal2011}.
As shown in the remaining panels of Figure \ref{fig_rho20_Qproperties}
the changes of the properties with galaxy density are small,
and the quasar properties depend only weakly on density.
However, due to the good statistics of our sample
the dependence on galaxy density is clearly detected
for some quasar properties
such as $M_i$, $L_\textrm{\s bol}$,$M_\textrm{\fns BH,vir}$, $R_\textrm{\fns FeII}$
and $\lambda$.
We find that luminosity ($M_i$, $L_\textrm{\s bol}$) decreases with galaxy density,
while black hole mass ($M_\textrm{\fns BH,vir}$, $R_\textrm{\fns FeII}$) increases.
With such dependence of luminosity and black hole mass,
Eddington ratio ($\lambda$)
decreases with galaxy density.
No significant change is detected for color ($\Delta(g-i)$).

\section{Discussion}
The best-fit linear function of quasar density-galaxy density relation 
has a slope of 0.618,
which means that quasar density changes slower than galaxy density does.
It is emphasized even more with the offset of the first point
from the linear relation in Figure \ref{fig_rho20_nQ}.
In under-dense region,
quasar incidence is higher than what is expected from
the trend in denser regions.
This seems consistent with literatures.
Even though the detail depends on how AGN are selected 
(for example in the optical, X-ray, infrared, or radio),
the general trend that the AGN fraction in over-dense regions is lower than 
that in under-dense regions
has been identified by a number of studies
\citep{best2004,martini-etal2002,martini-etal2006,martini-etal2007,arnold-etal2009,hwang-etal2012,shen-etal2007}.
These works found that AGN fraction in galaxy groups 
and poor-to-moderate richness galaxy clusters is higher
than that in rich clusters, and AGN fraction is higher in field than clusters.
It has been also confirmed that this difference is not 
simply caused by the morphological mix of galaxies between environments,
by examining AGN fractions in different environments
using only early-type galaxies \citep{arnold-etal2009,hwang-etal2012}.

As an analogous quantity to the AGN fraction,
we calculate the quasar-to-normal galaxy ratio,  
defined as number of quasars divided by number of CMASS galaxies
in a given bin of galaxy density.
To be fair and precise, 
we should prepare quasars and galaxies
from an identical parent sample
with the same luminosity cut applied.
However, it is challenging to prepare such samples from existing heterogeneous survey data.
Instead, as a quick analysis, we use the data we have used in this work.
Luminosity cut is separately applied to the two samples 
since luminosity is calculated at different K-correction redshifts.
We impose $M_i<-23.9$ to the quasar sample,
while we adopt two magnitude cuts, $M_i<-22.5$ and $M_i<-22.8$, for the galaxy sample.
The two magnitude cuts of the galaxy sample are considered 
to show the trend of the ratio does not change with the magnitude cut, 
thus compensating the arbitrary and inconsistent choice of luminosity thresholds.
\begin{figure}
\centering
\includegraphics[width=\columnwidth]{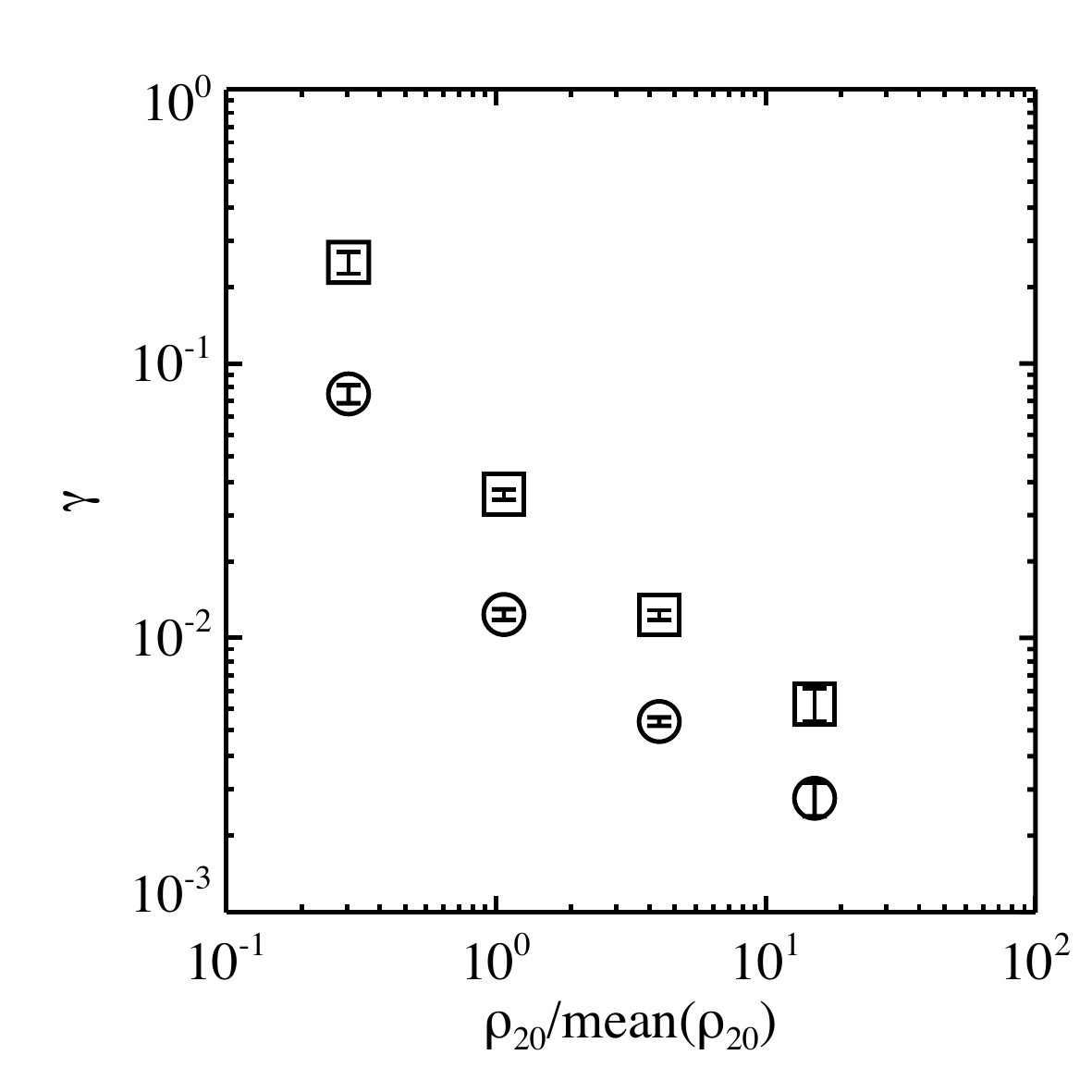}
\caption{Quasar-to-normal galaxy ratio ($\gamma$) as a function of galaxy density.
At a given galaxy density, number of the quasars of $M_i<-23.9$
divided by number of the CMASS galaxies of $M_i<-22.5$ (circle) or
of $M_i<-22.8$ (square).
Galaxy density bins are the same with one in Figure \ref{fig_rho20_nQ},
but the most under-dense region bin does not appear in this plot
since there is no CMASS galaxy in that bin by definition.
Error bars are calculated in the same way with the error bars in Figure \ref{fig_rho20_nQ}.}
\label{fig_rho20_Qfrac}
\end{figure}

Figure \ref{fig_rho20_Qfrac} shows that the quasar-to-normal galaxy ratio ($\gamma$)
decreases as one goes to denser regions.
Despite different sample criteria (quasars vs. AGN),
different physical quantities (ratio vs. fraction),
and different measures of environment 
(local galaxy density vs. clustercentric radius/velocity dispersion of a system)
the results of the literatures and us agree on
that AGN phenomenon seems to exhibit more efficiently 
in under-dense region than in over-dense region.
Galaxies in under-dense region tend to have 
richer cold gas reservoirs than those in over-dense region 
\citep{davies_lewis1973,solanes-etal2001,diSeregoAlighieri-etal2007,grossi-etal2009,cortese-etal2011,catinella-etal2013}.
In such circumstance, 
the central engine of AGN-host galaxies in under-dense region may be fueled more sufficiently 
and be more likely in quasar mode,
resulting in a larger fraction of galaxies being observed as quasars.
A similar trend was already shown by \citet{best2004} with lack of emission line AGN in clusters,
which is also explained well by lack of cold gas in clusters.

However, the anti-correlation between AGN fraction and environment
does not hold at higher redshift such as $z>1$ \citep{martini-etal2013}.
AGN fraction in clusters evolves so rapidly with redshift 
\citep{eastman-etal2007,galametz-etal2009,martini-etal2009}
that AGN fractions in the field and clusters become consistent at $1<z<1.5$.
Studies at higher redshifts like $z>2$ \citep{lehmer-etal2009,digby-north-etal2010}
even show a reversal of the anti-correlation.
Based on these studies, we can expect that the linear relation 
between quasar density and galaxy density we found 
may also evolve with redshift to have a steeper slope at a higher redshift.
It might imply that quasars can be used as an efficient probe 
of dense environment such as protoclusters at high redshifts.
\citet{orsi-etal2015} performed a numerical study of protoclusters
associated with radio galaxies and quasars at $2<z<6$,
and studied their connection to present-day cluster descendants.
By finding high-$z$ protoclusters with high-$z$ quasars,
the connection can be studied observationally.

The dependence of quasar properties on galaxy density seen 
in Figure \ref{fig_rho20_Qproperties} is understood 
in the frame of hierarchical merging scenario 
with taking into account the trend of cold gas reservoir of a galaxy
with environment aforementioned.
A galaxy in denser environment has experienced more mergers and interactions,
thus its central engine, supermassive black hole (SMBH), is heavier,
which is seen from the trend of $M_\textrm{\fns BH,vir}$ and $R_\textrm{\fns FeII}$.
Decreasing trend of the average quasar luminosity ($M_i$, $L_\textrm{\s bol}$) 
and activity ($\lambda$) with galaxy density
is due to the lack of fuel (cold gas) to feed the SMBH in denser environment,
even though quasars in denser region have deeper gravitational potential by more massive SMBH
to drag fuel more effectively.

Similar, but stronger dependence of properties 
such as color and star formation rate (SFR)
on environment have been observed in galaxies as well,
and the dependence reflect the amount of cold gas in different environments
\citep[e.g.,][]{hashimoto-etal1998,kauffmann-etal2004,skibba-etal2009}.
\citet{wijesinghe-etal2012} found that the SFR-density relation at $z<0.2$ is visible 
when both passive and star-forming galaxy populations are considered, 
while it is not seen when looking at the star-forming population only.
It might indicate that at that redshift range the primary effect of environment on galaxy properties 
comes from gas removal process rather than gas supply process.
Both star formation and AGN activities depend to the first order on the presence of cold gas.
One could therefore expect environmental effects acting on cold gas to produce similar environmental 
trends in both star-forming and AGN populations. However, the gas directly feeding star formation 
is distributed differently than the one powering the AGN (on larger scales vs. more centrally concentrated).
The SMBH feeding process is also expected to be stochastic 
\citep{hopkins_hernquist2006,peng2007,jahnke_maccio2011,hickox-etal2014}.
As a consequence, the fuel of the AGN will tend to be better shielded against
environmental effects than that of star formation,
which might be transferred into weak dependence of quasar properties on environment.

Meanwhile quasar clustering studies with CFs have reported weakly
positive correlation between quasar luminosity and host DMH mass 
\citep{adelberger_steidel2005,croom-etal2005,lidz-etal2006,coil-etal2007,
hopkins-etal2007,myers-etal2007,krumpe-etal2010,krumpe-etal2012,shen-etal2013}.
They found that more luminous quasars reside in marginally more massive DMHs,
which appears to reflect the opposite trend that we find here.
However, both effects are relatively weak, and the difference may be due to the fact that
we are taking the opposite approach to clustering studies for connecting 
quasar properties to large-scale environment.
Our results appear to follow the well-established trends
observed for the amount cold gas as a function of environment,
and so may provide a useful clue with regards to quasar fueling.

\section{Conclusion}
In this paper, we construct a galaxy number density field
using the SDSS DR12 CMASS catalog and study
how galaxy density affects the SDSS DR7 quasar properties.
The relation between quasar density and galaxy density
is well described with a linear function for logarithmic densities.
Quasars show weak dependence of their properties on environment:
SMBH mass increases with galaxy density, while luminosity decreases.
It should be noted that the linear relation
between quasar density and galaxy density
can make quasars a good tracer of LSSs
of the universe at high redshifts
where quasar are observed more easily than galaxies.
The relation makes it possible to trace the LSSs
in the galaxy distribution directly from observed quasars.
We do not detect any redshift dependence
of the relation in the narrow redshift range we use.
However, as discussed above, 
a redshift evolution of the relation 
between quasar density and galaxy density is expected --
quasars trace over-dense region more sensitively,
thus quasars can be a good marker of protoclusters at high redshift.
Moreover, the weak dependence of quasar properties on galaxy density
reinforces our argument that quasars can trace the LSSs
at different redshifts when the quasar properties might change systematically.
However, as mentioned before,
since the redshift range we consider in this study is narrow,
a further study with a wider redshift range is needed to confirm the weak dependence.

\acknowledgments
The authors thank the referee for thorough report.
It helped to improve the presentation of this work.
The authors thank Korea Institute for 
for Advanced Study for providing computing resources
(KIAS Center for Advanced Computation Linux Cluster System).
H. Lietzen acknowledges financial support from the Spanish Ministry 
of Economy and Competitiveness (MINECO) under the 2011 Severo Ochoa Program 
MINECO SEV-2011-0187.
M. Einasto is supported by the ETAG project
IUT26-2 of the Estonian Ministry of Education and Research and by the
project TK133, financed by the European Union through the European
Regional Development Fund.
H. Song thanks R. Gobat for his helpful comments on this work.

Funding for SDSS-III has been provided by the Alfred P. Sloan Foundation, 
the Participating Institutions, the National Science Foundation, 
and the U.S. Department of Energy Office of Science. 
The SDSS-III web site is http://www.sdss3.org/.

SDSS-III is managed by the Astrophysical Research Consortium 
for the Participating Institutions of the SDSS-III Collaboration 
including the University of Arizona, the Brazilian Participation Group, 
Brookhaven National Laboratory, Carnegie Mellon University, 
University of Florida, the French Participation Group, 
the German Participation Group, Harvard University, 
the Instituto de Astrofisica de Canarias, 
the Michigan State/Notre Dame/JINA Participation Group, 
Johns Hopkins University, Lawrence Berkeley National Laboratory, 
Max Planck Institute for Astrophysics, 
Max Planck Institute for Extraterrestrial Physics, 
New Mexico State University, New York University, Ohio State University, 
Pennsylvania State University, University of Portsmouth, 
Princeton University, the Spanish Participation Group, 
University of Tokyo, University of Utah, Vanderbilt University, 
University of Virginia, University of Washington, and Yale University.

\bibliographystyle{apj}
\bibliography{ms_rfrp1}{}

\begin{figure}[!h]
\centering
\includegraphics[width=0.6\columnwidth]{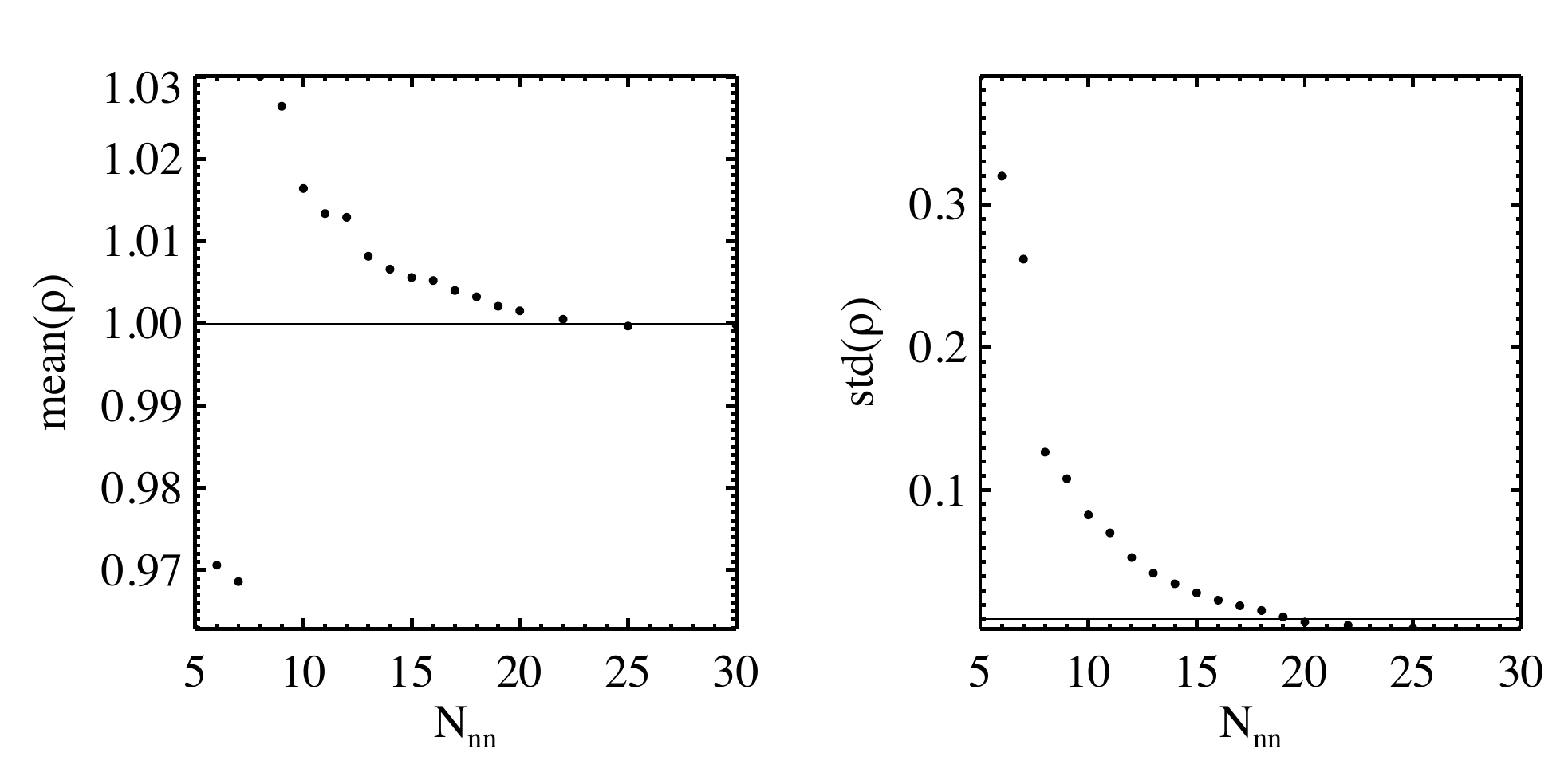}
\caption{
(Left) Mean of the densities measured at random spatial points 
by using the Spline kernel smoothing with the number of the neighbors within the kernel,
$N_\textrm{\fns nn}$, varied from 6 to 25.
The solid horizontal line represents the true density.
(Right) Standard deviation of densities as a function of $N_\textrm{\fns nn}$.
Solid horizontal line represents $1\%$ of the density.
}
\label{fig_ns_rhobar_std}
\end{figure}

\appendix
We perform a simple test to estimate the reliability of the smoothed density field when
$N_\textrm{\fns nn}=20$ is used.
We place particles at uniformly spaced mesh points to mimic a uniform density field,
and measure the density at random spatial points using 
the Spline kernel smoothing method for various values of $N_\textrm{\fns nn}$.
Figure \ref{fig_ns_rhobar_std} shows the mean and standard deviation of the measured densities
as a function of $N_\textrm{\fns nn}$.
As $N_\textrm{\fns nn}$ increases, the mean density approaches the true value 
(the solid horizontal line in the left panel of Figure \ref{fig_ns_rhobar_std}) 
with a smaller standard deviation.
$N_\textrm{\fns nn}=20$ is the smallest value 
at which the standard deviation drops below $1\%$.
This experiment shows that the error in the smoothed density measure 
falls below $1\%$ for $N_\textrm{\fns nn}\ge20$ 
in the case of the uniform particle distribution.
The error for clustered particle distributions depends on the ratio between 
the clustering scale and mean particle separation for each given $N_\textrm{\fns nn}$. 
For a clustered distribution, larger uncertainty than $1\%$ is expected, 
but we checked that even the $10\%$ uncertainty in the galaxy density estimation 
has less than $1\sigma$ effect on the 
relation between the quasar and galaxy densities found in the paper.

\end{document}